%% file: main.tex
\documentclass[sigconf,screen=true,anonymous=false,timestamp=false,acmthm=true,authordraft=false,authorversion=false,review=false]{acmart}

\usepackage{booktabs} 

\theoremstyle{definition}
\newtheorem{lesson}{Lesson Learned}

\copyrightyear{2018} 
\acmYear{2018} 
\setcopyright{rightsretained} 
\acmConference[PEARC '18]{Practice and Experience in Advanced Research Computing}{July 22--26, 2018}{Pittsburgh, PA, USA}
\acmBooktitle{PEARC '18: Practice and Experience in Advanced Research Computing, July 22--26, 2018, Pittsburgh, PA, USA}
\acmDOI{10.1145/3219104.3219164}
\acmISBN{978-1-4503-6446-1/18/07}

\begin{document}
\title{From Bare Metal to Virtual}
\subtitle{Lessons Learned when a Supercomputing Institute Deploys its First Cloud}

\author{Evan F. Bollig}
\affiliation{%
  \institution{Minnesota Supercomputing Institute\\University of Minnesota}
  \streetaddress{599 Walter Library\\110 Pleasant St SE }
  \city{Minneapolis} 
  \state{Minnesota} 
  \postcode{55455}
}
\email{bollig@umn.edu}

\author{James C. Wilgenbusch}
\affiliation{
  \institution{Minnesota Supercomputing Institute\\University of Minnesota}
  \streetaddress{599 Walter Library\\110 Pleasant St SE }
  \city{Minneapolis} 
  \state{Minnesota} 
  \postcode{55455}
}
\email{jwilgenb@umn.edu}

\newcommand{\note}{\footnote}

\begin{abstract}

As primary provider for research computing services at the University of Minnesota, the Minnesota Supercomputing Institute (MSI) has long been responsible for serving the needs of a user-base numbering in the thousands. 

In recent years, MSI---like many other HPC centers---has observed a growing need for self-service, on-demand, data-intensive research, as well as the emergence of many new controlled-access datasets for research purposes. 
In light of this, MSI constructed a new on-premise cloud service, named Stratus, which is architected from the ground up to easily satisfy data-use agreements and fill four gaps left by traditional HPC. The resulting OpenStack cloud, constructed from HPC-specific compute nodes and backed by Ceph storage, is designed to fully comply with controls set forth by the NIH Genomic Data Sharing Policy. 

Herein, we present twelve lessons learned during the ambitious sprint to take Stratus from inception and into production in less than 18 months. Important, and often overlooked, components of this timeline included the development of new leadership roles, staff and user training, and user support documentation. Along the way, the lessons learned extended well beyond the technical challenges often associated with acquiring, configuring, and maintaining large-scale systems.

\end{abstract}

%
%





\maketitle

\input{body}

\bibliographystyle{ACM-Reference-Format}
\bibliography{bibliography} 

\end{document}

%% file: body.tex
\section{Introduction}


Since the 1980's, large university facilities charged with supporting computationally intensive research have undergone some common changes to their cyberinfrastructure. Evidence of these changes is clearly reflected in the evolution of the names of these facilities.  For example, facilities opened in the 1980's typically include Supercomputing in their titles. Facilities begun in the 1990's often use words like High Performance Computing or HPC for short in their name. More recently, Research Computing seems to be included in the name du jour.  Importantly, these name changes reflect a change in focus from the cyberinfrastructure used to support the research, to the research supported by the cyberinfrastructure.

The Minnesota Supercomputing Institute (MSI) has been supporting computationally intensive research since 1984.  High-Performance Computing services are provided to approximately 700 different user groups representing both public and private entities on our campuses and throughout the state of Minnesota, with an active user base exceeding 4,000 annually. 
The Institute operates as an academic unit under the Office of the Vice President for Research. Its 42 full time staff belong to one of five functional groups. Three of these groups are focused on providing solutions for internal and external funded projects, while the other two groups fulfill general functions related to onboarding new users, accounts management, and supporting our data storage and computational infrastructure. On average, MSI recaptures less than half of its staffing costs from externally supported work. 

For many years, MSI strove to support the computational requirements of a fairly diverse user base by maintaining a relatively open and homogeneous platform. Infrastructure and workflows were generalized to simplify management and provide consistency in the user experience. 
In rare instances, ``edge-cases'' arose and small alterations to service building blocks could be made to accommodate most data-use agreements without greatly adding to Institute's systems monitoring and maintenance overhead.  For example, if an agency's data-use guidelines prohibited archiving data as part of a regular backup routine, then a separate volume could be created on our high performance file system that would not be part of a backup routine.

A short while ago, researchers across disciplines began to appear with increasingly large storage and computational requirements, as well as more stringent data-use policies. What were once ``edge-cases'' became cumbersome to manage and an increasingly large portion of our externally funded research portfolio. Compounding this issue was the fact that many of these groups lacked an alternative place--internal or external to the University--to conduct what was generally considered to be data-intensive research, but which didn't quite fit the mold for traditional HPC. 

A compelling case was made that MSI is well suited to host this new research as the primary provider for research computing services at the University. This in turn led to more intensive planning and discussions about what changes MSI would need to make to its existing cyberinfrastructure to accommodate and sustain these new demands. 

From these beginnings, MSI developed an ambitious timeline to deploy a scalable and sustainable system called Stratus. The new system was deployed within a year and entered production six months later. In this time, valuable lessons were learned that extended well beyond the technical challenges often associated with acquiring, configuring, and maintaining large scale systems. 


\section{Related Work}


Stratus is a subscription-based Infrastructure-as-a-Service for research on controlled-access data (a.k.a. protected data, restricted-use data, etc.) within a self-service environment. It was designed expressly to satisfy the NIH Genomic Data Sharing (GDS) Policy for the Database of Genotypes and Phenotypes (dbGaP) \cite{dbgap:2007}. Stratus is powered by the Newton version of the OpenStack cloud platform \cite{openstack:2017}, and is backed by the Luminous release of Ceph storage \cite{ceph:2006}. The integration, stability, scalability, and value of OpenStack and Ceph have been vetted by large deployments like CERN \cite{Cern:2015} and the NSF funded JetStream project \cite{Jetstream:2015}. 


Stratus is most similar in spirit to the Bionimbus Protected Data Cloud (PDC) \cite{Bionimbus:2014}; a fully GDS- and HIPAA-compliant platform operated at FISMA Moderate at the University of Chicago as part of the Open Science Data Cloud. Both Stratus and Bionimbus PDC are built on OpenStack and have Ceph storage. Stratus storage is entirely Ceph, in contrast to the current Bionimbus PDC, which has two object stores for protected data: 400 TB Ceph S3 and 1.7 PB IBM CleverSafe S3. The Bionimbus PDC cluster has 136 compute nodes for a total of 3168 vCPUs and 18 TB of RAM. At present, Stratus is a much denser system with 2240 vCPUs and 5 TB of RAM on only 20 compute nodes. The scale of Bionimbus PDC, as well as its FISMA qualification, have helped the project achieve NIH Trusted Partner status, which allows for a complete persistent clone of the dbGaP data to be maintained in storage. Meanwhile, Stratus presents users a multi-tiered storage environment with a ``dbGaP Cache'', for a scratch-like lifecycled object store where active subsets of the dbGaP data are cached for up to 60-days. When an object lifecycle is complete, the data is automatically purged from the dbGaP Cache; thereby avoiding bit rot, satisfying data-use requirements, and conserving storage costs. Under the Open Science Data Cloud, Bionimbus PDC does not charge for usage, but investigators must apply for project quota allocations. 

In the same vein of protected data clouds for dbGaP data, the Cancer Genome Collaboratory \cite{CancerCollab:2018} at the Ontario Institute for Cancer Research, is a Ceph-backed OpenStack cloud for cancer-related data. Collaboratory is designed for maximum storage capacity, which is used to persistently clone, among other things, dbGaP data dealing with cancer pulled from Bionimbus PDC. The Collaboratory currently has 2592 vCPUs, 18 TB of RAM, and over 7 PB of storage. In contrast to Stratus, where service is based on an annual base subscription with \`{a} la carte options, Collaboratory users are charged flat rate usage fees per vCPU hour and per GB hour of storage. 
%

The NIH has also sponsored three public cloud pilots for their dbGaP data: a) Broad Institute FireCloud \cite{FireCloud:2017}; b) Seven Bridges Cancer Genomics Cloud \cite{SevenBridges:2017}; and c) the Institute for Systems Biology Cancer Genomics Cloud \cite{ISBCGC:2017}. All three pilots run on public cloud providers (Google/Amazon), where the annual cost of computing is substantially higher than on a local on-premise cloud. To alleviate this burden, each of the pilots is offering new investigators \$300 in credits to get started, and grant opportunities of up to \$10,000 per project. 

%


\section{Background}

\begin{figure}[!t]
\centering
\includegraphics[width=\columnwidth]{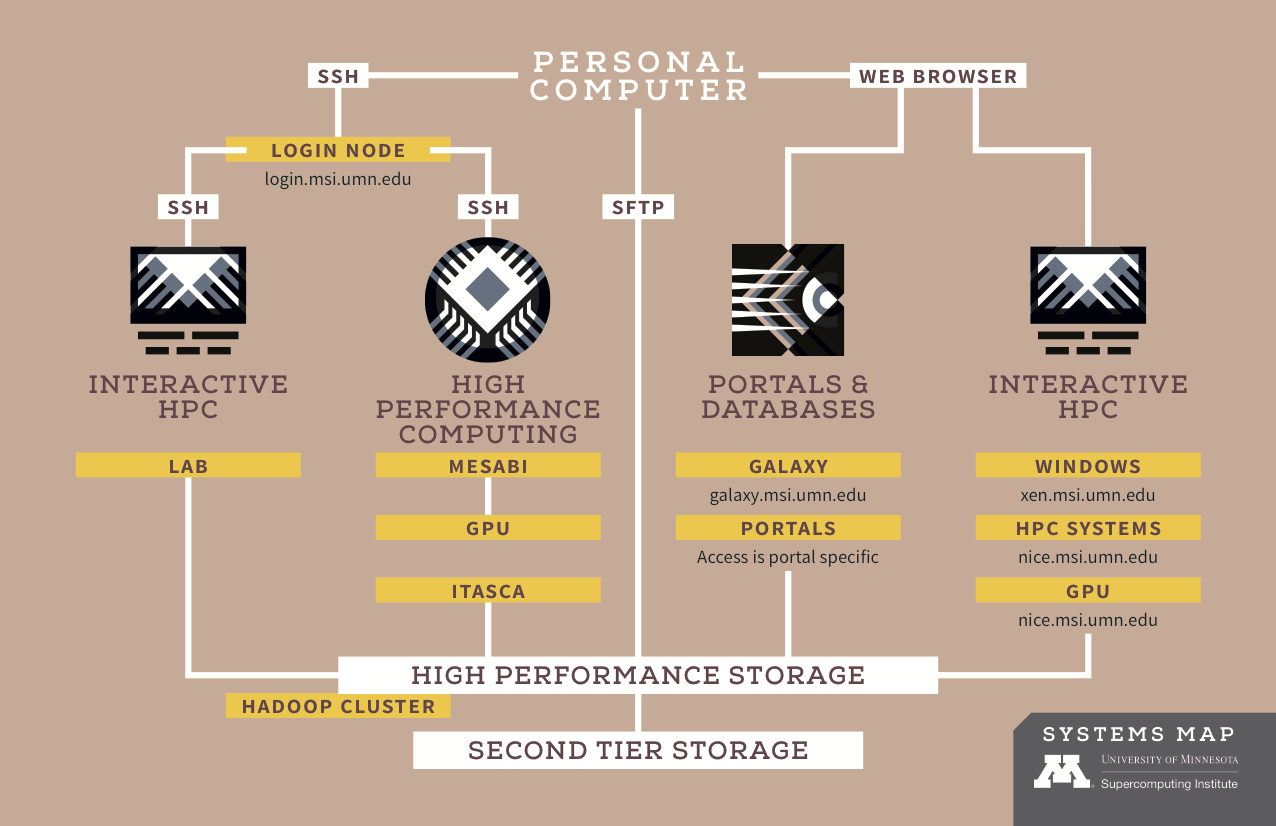}
\caption{MSI core-service portfolio. }
\label{fig:msi_services}
\end{figure}

The Institute's core-service portfolio is shown in Figure~\ref{fig:msi_services}. At the center, MSI's focal point is traditional HPC clusters for batch-scheduled jobs. The batch experience extends behind web portals and workflow managers (e.g., Galaxy \cite{afgan2016galaxy} and JupyterHub \cite{Milligan:2017}) where job workflow and submission are abstracted with web interfaces. In addition to batch, interactive computing resources and remote desktop solutions are first-tier services at MSI, but focus predominantly on workflow prototyping, debugging code, and data visualization.

MSI's flagship cluster, Mesabi, is homogenous in CPUs (Intel 64-bit Haswell), with some heterogeneity in memory (e.g., subsystems composed of 64 GB, 256 GB, and 1 TB nodes). Mesabi has a 40 node GPU subsystem with a total of 80 NVidia K40s, as well as 40 nodes with solid state drives. Although the system was installed in 2015, Mesabi is still among the top 20 university-owned supercomputers in the nation, with a peak performance of over 670 Tflops and an additional 105 Tflops from the GPU subsystem. 

In recent years, the Institute made a dedicated push to homogenize the user experience across services with the philosophy that research computing should be simple and accessible. For example, attached to all compute services is a 4 PB global-namespace high-performance Tier I storage for hot data and user home directories---this ensures that wherever a user computes, their data is immediately within reach. Similarly, Second Tier Storage (a.k.a. Tier II storage) is a globally accessible S3 object store for cool data. On top of all core-services, MSI also integrates the University's central services for authentication and identity management.  

Holistically, the homogenous environment at the Institute satisfies the majority of MSI's more than 4,000 user workflows, and fits most data-use agreements. For infrequent edge-cases that are not an obvious fit, one-off concessions or tweaks to services are often possible. What motivated the Stratus project, however, was the realization that some edge-cases could not be satisfied, and other cases were quickly becoming untenable for time-limited staff, especially due to complexity and quantity. 

\section{Requirements \& Planning}

From inception, the goal of the Stratus project was to deploy a compute cloud to complement the Institute's existing service portfolio. The new service was not intended to compete with, or supplant, existing HPC resources---though HPC-like performance was required. Here we present the four driving requirements, and the proposed solution that became Stratus. 

\subsection{Controlled-Access Data Enclave} 

The primary need for Stratus was an enclave for research on controlled-access data demanding large amounts of storage and compute for analysis. 
The principal example of data demanding this environment is the NIH Database for Genotype and Phenotype (dbGaP) \cite{dbgap:2007}, which is currently on the order of hundreds of TBs of data and growing. dbGaP is actively used by as many as 40 groups at the University, with interest increasing annually. 

dbGaP data is classified as either open- or controlled-access. Open-access data can be freely accessed and processed on any system at the University, whereas controlled-access data is constrained to users with an appropriate data-use agreement, and must be processed on resources that meet an adequate number of controls. Examples of such controls include disabled backups for the data storage, two-factor authentication, access logging, etc. The nature of the data is sensitive and therefore necessitates moderate security as prescribed by the NIH Best Practices Guide \cite{dbgapBestPractices:2015}.  

Though other entities at the University manage access-controlled data---e.g., the University's Academic Health Center is the primary covered entity for HIPAA data, the Office of Information Technology handles ITAR data, etc.---the intended purpose of dbGaP is for research and does not align well with the mission of those entities. Likewise, the sheer size of dbGaP data would be a challenge for the expertise, capacities, and compute resources around the University, but fits easily within the wheelhouse of MSI. Thus, a compelling case was made for MSI to accommodate dbGaP data and prepare for other emerging controlled-access datasets in the future. 

Early efforts to accommodate dbGaP on MSI core-services exposed difficulties in handling the data with one-off modifications. For example, on global Tier I storage, a dedicated ``single\_copy'' volume was created with data snapshots disabled. However, since the volume is part of global-namespace storage, it presents a huge potential for data leakage. Users could inadvertently copy dbGaP data onto another volume where the data would roll into an automatic snapshot. Cleaning up such an incident would necessarily trigger a purge of Tier I snapshots for compliance, to the detriment of users who depend on snapshots to exist. In similar fashion, file permissions could be recklessly opened, leaving data visible to any number of unauthorized users across services. In the end, MSI put substantial staff effort into user trainings to combat such scenarios, but the solution was ultimately seen as a short-term stop-gap. Managing dbGaP on existing core-services was too heavy-weight and unscalable, and the Institute opted to invest in a new environment where the controlled-access data experience is integrated into the design. 

\subsection{On-Demand Resources} 
The Institute also recognized the growing need for other on-demand data-intensive (i.e., ``big data'') research. 
Although capacity exists for these cases on the global Tier I filesystem and Mesabi, the ultimate challenge has always been meeting requests on-demand. It can often be difficult to reserve more than a couple nodes for interactive use on Mesabi, even during off-peak hours. Furthermore, this type of research often depends on non-traditional HPC software and interactive workflows that do not integrate well with multi-user, scheduled resources. It is also worth noting that this gap requires resources with higher availability of services, as well as a self-service mode of operation. 

\subsection{Long-Running Jobs} 
MSI implements an operational plan for core services that includes a mandatory monthly maintenance window. On the first Wednesday of every month, the Institute performs routine updates, security patches, and necessary reboots on infrastructure. To avoid data loss, jobs are held in queues when impacted by the scheduled outages. Consequently, batch jobs can have a maximum wall-time of 29 days, and jobs near the limit can only run immediately following scheduled downtimes. Although solutions exist to checkpoint and restart software, it is not always possible to integrate such technology into black-box software, and thus a need still exists for MSI to support continuous operation beyond the 29-day mark. 

\subsection{Container-Based Computing}
Circa 2013, the Institute investigated the implications of integrating Docker containers \cite{docker:2014} into HPC workflows. At the time, concerns with the security of containers prevented adoption of the technology on bare-metal HPC systems---it was decided that containers could only be trusted if run on virtualized hardware. During the intervening years, container-security has improved, and interest in the scientific community has grown. Stratus was seen as an opportunity to embrace the new technology so long as containers run in an isolated sandbox.




%

%


\subsection{Proposed Solution}

\begin{figure}[!t]
\centering
\includegraphics[width=\columnwidth]{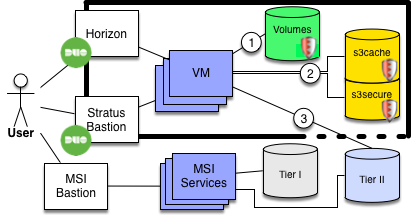}
\caption{Stratus as a walled-garden, independent of other MSI services.}
\label{fig:stratus_sandbox}
\end{figure}

The resulting design of Stratus is shown in Figure~\ref{fig:stratus_sandbox}. At a high level, Stratus is an on-premise cloud environment for self-service research computing. The new service is a standalone ``walled garden'' and is isolated from the rest of MSI services---as denoted by the heavy black border. Additionally, Stratus was designed following the details of the NIH GDS Best Practices Guide \cite{dbgapBestPractices:2015} as required controls.

Two ports of entry exist into Stratus: a) a web-interface called Horizon through which users can self-manage virtual machines, storage, etc.; and b) the Stratus Bastion terminal server for secure shell access into the interior. Both paths into Stratus require Duo \cite{Duo:2017} two-factor authentication against the University's central identity management system. Inside the environment, users are allowed to create Virtual Machines (VMs), self-manage software---including containers---and attach/detach storage as needed for their research. 

Three tiers of storage are accessible in Stratus:
1) Volumes suitable for active/hot data for open- and controlled-access datasets; 2) Two types of object storage for cool data that remain controlled-access; and 3) External access to MSI's Tier II storage for open-access data---this includes distilled derivatives that qualify for open-access. Shields denote storage suitable for controlled-access dbGaP data. With respect to 2), the first object storage is a ``dbGaP Cache'' (S3Cache), or short-term storage intended for dbGaP reference data. The Cache not only has capacity for the large datasets, but also a life-cycle policy to purge data periodically. The second object store (S3Secure) matches the Cache in everything except lifecycle, and thus allows for more persistent storage of data. Note that on all tiers, users must abide by their data-use agreements and purge data when necessary. 


%

\subsection{Timeline}

\begin{figure}[!t]
\centering
\includegraphics[width=\columnwidth]{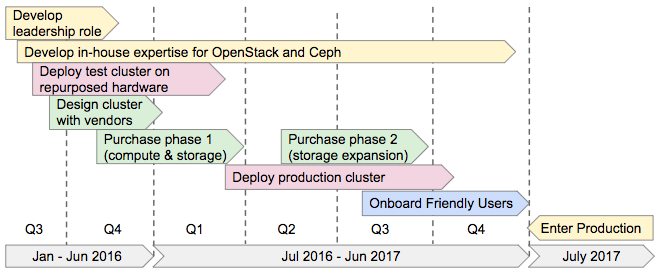}
\caption{The timeline to get Stratus into production.}
\label{fig:stratus_timeline}
\end{figure}


The final timeline to production is shown in Figure~\ref{fig:stratus_timeline}. 
Yellow milestones dealt with staff training and changing culture. Cloud computing is a substantially different experience compared to HPC computing, and users have more freedom to make decisions. Managing a cloud environment requires patience and preparedness from operators. Therefore, through the full course of the project, MSI invested in developing in-house expertise for cloud operations, use and management. Red milestones capture hardware deployments. Two clusters were built for Stratus: a) a trial cluster built on repurposed hardware to confirm the design of Stratus would function; and b) the production system on newly acquired hardware. Note that planning and purchasing the production system is a third set of milestones in green. Lastly, the blue milestone saw friendly users on-boarded to vet the system in the lead-up to the production state.  

\section{Training Staff and Changing Culture}


With the decision to invest in a cloud environment, it was clear that moving the organization's staff from their traditional base of support required some cultural change. 

Shortly before the project was initiated, MSI restructured as a matrix organization with five well-defined functional groups as verticals, and projects as horizontals. The new structure also defined two roles to clarify reporting lines and priority: a) a Project Manager to oversee the day-to-day movements of the project through to operational hand-off; and b) the Project Staff role for anyone allocated to the project. Project Staff report first to their functional-group lead to ensure that intra-group efforts and emergencies take priority. Beyond orchestrating Project Staff efforts, the Project Manager works closely with functional leads to communicate inter-group efforts, adjust staff allocations, and keep the project moving forward.

Stratus had 7 staff allocated the project. In many cases, staff efforts overlapped, but the broad categories of effort and estimated percentages of Full Time Employment (FTE) were: a) Project management (30\%); b) OpenStack deployment and development (70\%); c) Ceph deployment (40\%); d) Acceptance tests and benchmarks (25\%); e) System Security (10\%); and f) Networking (10\%).




\lesson{Take ownership of the project. Lead by example.} 

When early plans for Stratus were first introduced, some staff vehemently opposed the idea, arguing that self-managed VMs would undermine the security of the Institute's entire infrastructure. Empowering 
users to manage their own VMs was seen as the inverse of the fully-managed HPC philosophy that MSI was founded on. Questions also arose around the choice of MSI, not another organization, to shoulder this research-centric service. Lastly, concerns were voiced about managing controlled-access data in an Institute that had a long history of denying requests for users to work with protected data.  These were very clearly areas of cultural resistance that MSI had to hash through.

The project manager---with the backing of functional group leads---held a number of ``therapy sessions'' (meetings) to present cloud technologies, dispel myths, and open a dialog about MSI's evolving role to serve the research computing community. Sessions helped calm fears and demystify ways that an organization like MSI prepares for protected data. Periodic reminders were also necessary to keep staff focused on dbGaP controls and avoid generalizing to the myriad of other data types and compliances; those could be added incrementally later. 


 
\lesson{Staff do not always appreciate new services (a.k.a. ``responsibilities''). Expect pushback.}

With the Institute staff numbering 42, the available capacity to tackle projects is limited. 
Any reluctance or resistance could set back the entire project. Although the Institute's matrix structure allowed for management to select staff for the project, volunteers were sought with the understanding that they would invest their energy more efficiently, own problems, and mesh better in teams. 

MSI highlighted professional development opportunities to gather volunteers. First, research experience and co-authorship. While it may seem insignificant to many in science, system admins are often unsung heroes facilitating research with little or no recognition. Thus far, four staff have proudly become first-time co-authors on the Stratus project (\cite{Stratus:2017}). Second, MSI emphasized r\'{e}sum\'{e} building by cross-training staff in storage, networking, automation, etc. The most compelling opportunity was the possibility to develop skills with Ceph and OpenStack as operators and administrators; in so doing, the Institute was investing in skills touted to make its employees 36\% more valuable in industry \cite{openstack_value:2013}. 
Volunteers flocked to the project, and much of the success is due to their in-kind investment back into MSI.

\section{System Configuration, Acquisition, and Installation}

Before the Stratus project kickoff, the choice to use 
OpenStack \cite{openstack:2017} and Ceph \cite{ceph:2006} had already been made. 
OpenStack was selected for multiple reasons: a) it has built-in support for multi-tenancy, software defined networking, logging, SSL/TLS encrypted traffic and other features to make compliance easier; b) the modular OpenStack services enable mix-and-match configuration and are decoupled for fault tolerance; and c) a massive open source community, including groups from the NSF JetStream project, CERN, and NASA distinguish the project as the leading software for cloud deployments across both industry and scientific computing. 
Similarly, Ceph is free, open source, and backed by much of the same community as OpenStack. Ceph excels as an efficient, low-cost storage platform with the ability to scale to many PBs. MSI had already deployed an instance of Ceph for its S3 Second Tier Storage for core-services, and the stability, scalability, and success of that service motivated its reuse for Stratus.

Given the choice of products, the process to select hardware configurations and complete an acquisition was executed in three stages: a) discovery, b) selection, and c) purchase. During the discovery phase, MSI requested from vendors examples of: a) fully-integrated OpenStack solutions (with any storage backing); b) reference architectures for OpenStack-only with a separate backing; and c) reference architectures for Ceph alone. Concurrently, project staff also began test deployments of both softwares on repurposed hardware from the MSI machine room to assess features like erasure coding on storage and live-migration of VMs. 



\lesson{Vendor solutions were rigid, expensive, and would not support HPC.}

The reference solutions provided by vendors were lackluster. Vendors promoted OpenStack as a featured product, but it was clear that their use-cases were not focused on HPC; more typically, configurations were for enterprise web-hosting with heavy over-subscription (e.g., 18x) and relatively small storage footprints (e.g., 16 TB). The fully-integrated solutions were rigid and did not allow for customizations. Further, popularity combined with the complexity of supporting OpenStack resulted in prohibitively high prices compared to the value of hardware.



\begin{figure}[!t]
\centering
\includegraphics[width=0.85\columnwidth]{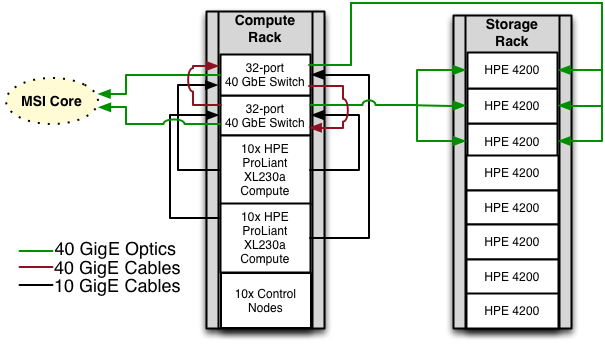}
\caption{The Stratus hardware composed of HPE compute and storage nodes.}
\label{fig:stratus_hardware}
\end{figure}

Based on benchmarking data from the development cluster, and encouraged by success in deploying OpenStack in-house, MSI saw an opportunity to use a slightly modified reference Ceph architecture from HPE for storage, and acquire HPC-specific hardware for OpenStack. The resulting architecture of Stratus is shown in Figure~\ref{fig:stratus_hardware} where 20 compute nodes, each a Mesabi 256 GB node (HPE XL230a), are attached to HPE 4200 high-density storage servers via two 40 GbE switches. Each HPE 4200 server contains 24 8 TB hard drives, six 960 GB SSDs and two 800 GB NVMe flash cards in 2U of rack space. 

\section{Security Planning}
Stratus is operated as a shared-responsibility security model similar to most cloud providers. Traditional cloud providers absolve themselves from responsibility of protecting users, focusing instead on achieving compliance within the underlying cloud hardware and leaving users to individually manage and vet whatever they run on top of the cloud; a potentially dangerous situation for novice users with protected data. Stratus was built with an MSI-first mentality, prioritizing security of the hardware, OpenStack, and Ceph. However, a fair amount of effort also went into sandboxing user VMs and providing sane defaults for security that users can opt-out of at their own risk. For example, Stratus only allows campus network traffic on ports and 443, and 8443 with SSL-encryption required, and projects cannot connect to VMs in other projects. 
\begin{figure}[!t]
\centering
\includegraphics[width=\columnwidth]{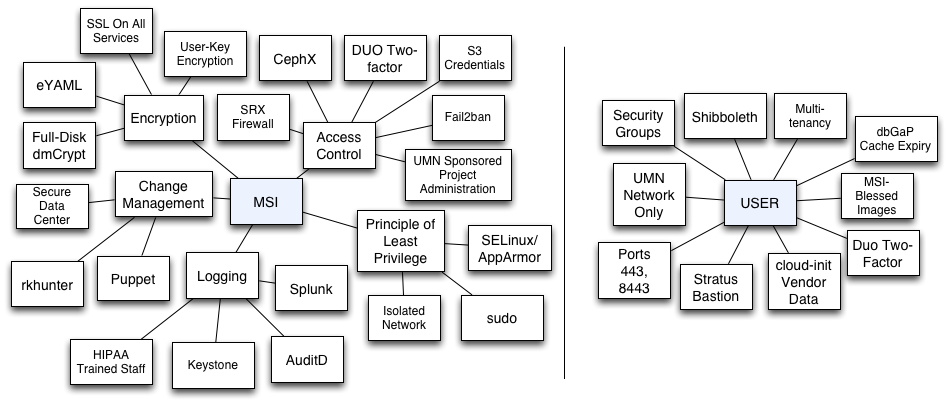}
\caption{Although Stratus has a shared-responsibility security model, effort was made to establish sane defaults for security within the user's space.}
\label{fig:stratus_security_model}
\end{figure}

Figure~\ref{fig:stratus_security_model} summarizes a number of efforts made by MSI to protect itself and users. 


\lesson{The NIH GDS Policy is a good launching point.} 

MSI used the NIH dbGaP Best-Practices Guide \cite{dbgapBestPractices:2015} as a checklist assuming every control to be required. This may seem excessive, but the controls for dbGaP data are fairly lax, and were easily met technically with OpenStack and Ceph. Also, the assumption was that using the additional upfront effort would simplify efforts later if Stratus was hardened to manage more stringent data-use agreements later. 



\section{Systems Installation and Testing}

Deploying the new system required a great deal of testing before transition to production. In part, this testing also presented an opportunity for staff to better understand nuances of the cloud experience, as well as limits of the system before users were dependent on it.  Likewise, friendly users were invited to test the system and report back on any shortcomings. 



\lesson{Network, Compute, Storage. Everything is present right? dbGaP will require more.}

Over the course of six months, MSI received all hardware, put in the necessary effort to install OpenStack and Ceph, and built the user experience. The assumption was that the cluster was complete, and the service ready to roll. Therefore, friendly users with dbGaP data were invited to take Stratus for a test-drive. One such user appeared with 120 TB of dbGaP data prepared to move into the cluster. The Institute discovered that the footprint of this user's data was not typical, but it was necessary to assume two or three heavy users of this magnitude would be active on the cluster at one time. Based on this feedback, the dbGaP Cache needed about 400 TB of object storage capacity (1 PB raw) capacity---the initial purchase of Stratus only accounted for approximately 100 TB of usable storage. Consequently, MSI was forced to delay the project timeline while an additional 5 nodes of storage were acquired, installed and testing could resume.  

\lesson{Even staff expect a managed HPC environment with pre-installed software.}

Training staff to use Stratus for testing and benchmarking took significantly more effort than anticipated. The first hurdle for staff was to understand the entirely new vocabulary of cloud. For example, the differences between a CPU-core and a vCPU, or the meaning of Block versus Object storage. When asked to benchmark a VM, the pain point became the absence of a queuing system and software libraries/modules that allowed their benchmark code to run. Such hurdles were obviously going to arise with regular users as well, so the lessons were well received. 
Ultimately, this was a nice reminder for staff to appreciate the unsung hero system administrators who manage the HPC systems and provide the resources they and users take for granted.

\lesson{Virtualization is not the antithesis of bare-metal.}

Stratus was designed to provide HPC-like performance, and MSI was pleased to see the design succeed. 
 On virtualized, but non-oversubscribed hardware,
HPL CPU benchmarks showed only a 5\% efficiency loss in Stratus VMs versus bare-metal Mesabi. Likewise, FIO benchmarks on both systems show Stratus' Volume storage capable of achieving up to 12.5\% more write bandwidth compared to the high performance filesystem on Mesabi. Obviously, there are major scaling and architectural differences between the two systems, but comparable performance between Cloud and HPC was a goal of the project, and Stratus certainly contends. Furthermore, the value of virtualization in terms of live-migration, snapshots, and on-demand resources makes the performance difference tolerable.


%

\section{Cost Recovery}

Given the specialized nature of Stratus, and the episodic nature of on-demand cloud utilization, it was decided that access would be limited to a paid subscription service in contrast to the free access MSI provides on most other services. This decision offers three benefits: a) attaching a cost creates enough of a hurdle 
to cull superficial users; b) recovered costs can be rolled into the acquisition of new hardware to expand the system; and c) users are presented a quantifiable penalty for allocated resources, motivating some (not all) users to think consciously about utilization and clean-up whenever possible (a.k.a., exercise good cloud hygiene).

The Institute opted for a zero-profit model for cost-recovery. Rates are calculated for each of Stratus' core components: a) vCPU Compute; b) Block Storage; and c) Object Storage. To calculate vCPU rates, the cost of all compute hardware is aggregated, plus the total staff FTE costs for support on the compute portion of the system (e.g., system administration, ticket triage, training, etc.). This number---the total cost of ownership for Compute---is divided by the number of CPU-hours possible over the lifetime of the hardware, and again by the oversubscription rate to get the cost per vCPU/hr. A 5-year lifecycle on hardware is assumed, and MSI targets 100\% recovery of the hardware when 85\% utilized to ensure that expansions happen before the system is critically full. This process is repeated to get a per TB/yr rate on each type of storage.

\lesson{Private clouds are significantly cheaper.

Table~\ref{tbl:stratus_rates} presents the FY2018 estimates for Stratus internal rates\footnote{Available only to the University of Minnesota. Rates are subject to change annually based on hardware and staff changes.}. To simplify enrollment, the Institute requires that projects subscribe to a minimum base package of 16 vCPUs, 32 GB RAM, and 2 TB of storage. Beyond the base subscription, \`{a} la carte pricing allows for additional vCPUs, TBs of block storage, and TBs of object storage. Memory is granted at a ratio of 2 GB per vCPU purchased. dbGaP users get free access to the dbGaP Cache. 

\begin{table}[!t]
\centering
  \caption{The Stratus cost-recovery fees for FY2018.}
  \label{tbl:stratus_rates}
    \begin{tabular}{ | l | l | l |}
    \hline
    Service Name & Unit & Cost/Year  \\ \hline\hline
    Stratus base subscription & Pkg & \$626.06 \\ \hline
    Additional CPU Cores & vCPU & \$20.13 \\ \hline
    Additional Block Storage & TB & \$151.95 \\ \hline
    Persistent Secure Object Storage & TB & \$70.35 \\ \hline
    \end{tabular}
\end{table}

To illustrate the value of private cloud, a comparison is made of the Stratus base subscription to similar resources on Amazon Web Services. It is assumed that a c5.4xlarge (16-vCPU, 32 GB RAM) AWS instance is comparable with 2 TB attached EBS GP2 Volume Storage. The cost on AWS is minimized by reserving 1 year and paying upfront for the reservation. At the current market value \cite{AWSCalc:2018}, AWS would cost approximately \$7337/yr or 11.7x higher than Stratus.



%
%
%

\section{Onboarding Users}

Training is mandatory to get into the Stratus environment. 
MSI has already developed three tutorials to onboard users and effectively use the new system for different use-cases and levels of understanding when it comes to system administration and software management. 

%

%
%
%
%

\lesson{Users do not truly understand what they are asking for.}

As the number of projects subscribing to Stratus grows, a recurring scene has played out during on-boarding tutorials. Generally, most users are excited to get to the cloud because they heard about the wealth of opportunity and flexibility promised by self-service and on-demand. In stark contrast to their expectations, however, the most popular questions asked during training revolve around: a) ``How do I run jobs?''; b) ``What happened to my data and software? It was in my home directory.''; and c) ``Where do I send requests for software installs, or system administration?''.  Clearly, users have grown accustomed to the Institute's fully-managed environment, globally mounted storage, and the ability to lean on technical staff to assist them at each stage of their work. 

In response to these questions, the on-boarding tutorial is now delivered incrementally with extra points of repetition to impress on users the novelty (good and bad) of operating in the Cloud. Patience is required for reality to set in that Cloud may not be the solution they are looking for. 


\lesson{Convenience trumps cost, and users will pay for POSIX.}
 
On the same day Stratus went into production (July 1, 2017), the first group to subscribe to Stratus requested 20 TB of Volume storage, or 10\% of the cluster's total usable capacity. With over 30 more dbGaP approved groups on campus, the service could have been sold faster than it could scale (even with cost recovery, there are limits to how fast purchase orders can move). The lesson was clear: Volume storage, with the ability to mount as a POSIX filesystem, is easy to work with, and the cost is low enough that groups will gladly pay the premium for convenience. MSI has since ramped up efforts to demystify S3 utilization for users and emphasize the benefits the free dbGaP Cache. Also, MSI is actively testing new tools to give users a POSIX-like experience when working S3 storage. 

\section{Future Directions}

Now that Stratus is live for dbGaP users, two  final lessons are apparent. 

\lesson{dbGaP is just the beginning.} 

The next extensions to compliance in Stratus could be FISMA or HIPAA. While these are ``desired'' by users, the ``need'' for either is much harder to establish. For most requests thus far, the solution to the problem is not a HIPAA compliant service, but rather a modified workflow that de-identifies data or distills it to a state that can be processed on MSI services, and then re-integrates the results once it leaves MSI. However, it is worth noting that new datatypes have emerged (e.g., whole-genomes), which now pose the challenge: how does one protect data when the data itself is the identifying information?

\lesson{Everyone wants a cloud. \label{lesson:last}} 

More and more use-cases are appearing in the user community in need of a flexible on-demand environment for computing. 
MSI is currently exploring ways to permit some of these additional customers access to Stratus without greatly expanding the infrastructure and without jeopardizing the security of the system. As of February 2018, a new General project space is available by request. The new class of projects get more freedom to open ports, but lose access to dbGaP Cache, and are otherwise the same self-service experience.

%


%
%
%
%
%
%
%
%
%
%

\section{Conclusion}
In conclusion, this document has presented \ref*{lesson:last}
 lessons learned during the design, acquisition and operation of an OpenStack Compute Cloud for the Minnesota Supercomputing Institute. These lessons should assist similar institutions which seek to deploy a local cloud, or to manage NIH dbGaP data.